\documentclass[10pt,letterpaper,superscriptaddress]{article}
\usepackage{opex3}
\usepackage{graphicx}
\usepackage{color}
\def\PRL{{Phys. Rev. Lett.} }
\def\PRA{{Phys. Rev.} A }
\begin{document}
\title{Testing Ultrafast Two-Photon Spectral Amplitudes via Optical Fibres}

\author{G.~Brida$^1$, V.~Caricato$^{1,2}$, M.~V.~Chekhova$^{1,3}$, M.~Genovese$^1$, M.~Gramegna$^1$, T.~Sh.~Iskhakov$^3$}
\address{$^1$Istituto Nazionale di Ricerca Metrologica, Strada delle
Cacce 91, 10135 Torino, Italy \\
$^2$Dipartimento di Fisica, Politecnico di Torino, Corso Duca degli Abruzzi 24, \\10129 Torino, Italy\\
$^3$Department of Physics, M.~V.~Lomonosov Moscow State University,
Leninskie Gory, 119992 Moscow, Russia} \email{m.genovese@inrim.it}

\begin{abstract}
We test two-dimensional TPSA of biphoton light emitted via ultrafast
spontaneous parametric down-conversion (SPDC) using the effect of
group-velocity dispersion in optical fibres. Further, we apply this
technique to demonstrate the engineering of biphoton spectral
properties by acting on the pump pulse shape.
\end{abstract}
\ocis{(270.0270) Quantum optics, (270.5585) Quantum information and
processing}

The spectral properties of two-photon light are fully
characterized
by two-photon spectral amplitude (TPSA). It was first introduced for
describing two-photon light generated via spontaneous parametric
down-conversion (SPDC) from femtosecond pulses and the corresponding
interference experiments~\cite{Grice,Keller}. Recently, TPSA has
been shown to determine the degree of frequency
entanglement~\cite{Fedorov}. In particular, the technique of
heralded generation of pure single-photon quantum states relies on
testing TPSA~\cite{Mosley}.

Experimental measurement of TPSA has been first performed by
registering the distribution of coincidences between signal and
idler photons as a function of frequencies selected in the signal
and idler channel (joint spectral
distribution)~\cite{Kim&Grice,Wasilewsky,Poh}. For biphotons
generated from a cw pump, when TPSA depends on only a single
frequency argument, an alternative technique has been suggested for
its measurement, based on its spreading in optical
fibres~\cite{spectronPRL,spectron}. This technique, in particular,
enabled observation of various Bell states contained within the
width of the TPSA~\cite{BellPRL}.

In the case of SPDC pumped by short pulses, TPSA depends on both
signal and idler frequencies, as signal and idler photons are not
any more delta-correlated in frequency. Propagation of pulsed
two-photon light through optical fibre has been studied in
Ref.~\cite{Silberhorn}, and it was suggested to use dispersion
spreading for the spectroscopic study of conditionally prepared
single-photon states. In an earlier paper~\cite{Kim}, a similar
technique was used for the study of single-photon states prepared
via cw SPDC.

In this paper, we show that propagation of two-photon states through
a dispersive medium (for instance, an optical fibre) provides more
general information. Namely, it provides a frequency-to-time
two-dimensional (2D) Fourier transform of the TPSA and this enables
one to characterize TPSA by measuring time intervals. Further, we
apply  this method to a proof-of-principle demonstration of the
engineering of spectral properties of biphoton entangled states. A
very efficient technique of measuring two-dimensional coincidence
time distribution has been suggested in Ref.~\cite{Silberhorn}.
Here, we use a simpler method,  based partly on time selection and
partly on frequency selection. Despite its simplicity, our method
allows us to observe such effects of pulsed TPSA as asymmetry with
respect to signal-idler exchange and interference structure.

For SPDC from a short-pulsed pump, the two-photon state has the
form~\cite{Grice,Keller}
\begin{equation}
|\Psi\rangle=\int\int\hbox{d}\omega_s\hbox{d}\omega_i
F(\omega_s,\omega_i)a^{\dagger}(\omega_s)a^{\dagger}(\omega_i)|\hbox{vac}
\rangle, \label{state}
\end{equation}
where $a^{\dagger}(\omega_s),a^{\dagger}(\omega_i)$ are creation
operators of the signal and idler photons. The TPSA
$F(\omega_s,\omega_i)$ depends both on the pump spectrum and on the
phase matching conditions in the crystal~\cite{Grice}. It is
convenient to introduce the deviations of the signal and idler
frequencies $\omega_s,\omega_i$ from exact phase matching, which,
for simplicity, we will consider as frequency-degenerate at
frequency $\omega_0$: $\omega_s=\omega_0+\Omega_s$,
$\omega_i=\omega_0+\Omega_i$. The two-photon time amplitude (TPTA),
whose physical meaning is the probability amplitude of the signal
photon registered at time $t_s$ and the idler photon at time $t_i$,
is the 2D Fourier transform of TPSA:
\begin{equation}
F(t_s,t_i)\propto\int\int\hbox{d}\Omega_s\hbox{d}\Omega_i
e^{i\Omega_s t_s}e^{i\Omega_i t_i}F(\Omega_s,\Omega_i). \label{TTPA}
\end{equation}

In a dispersive medium, as, for instance, an optical fibre of length
$l$, each of the photon creation operators from Eq.~(\ref{state})
acquires a frequency-dependent phase~\cite{spectron} that can be
attributed to the TPSA. As a result, the TPSA changes:
\begin{equation}
F(\Omega_s,\Omega_i)\rightarrow F(\Omega_s,\Omega_i)e^{i
l(k''_s\Omega_s^2+k''_i\Omega_i^2)/2}, \label{fibre}
\end{equation}
where $k''_{s,i}$ are second-order derivatives of the dispersion law
for signal and idler photons. Transformation (\ref{fibre}), as is
well-known in the diffraction theory or in the theory of dispersive
spreading of short pulses, at a sufficiently large length $l$
(`far-field zone') leads to a Fourier transformation from frequency
to time. Note that this transformation will occur in both $\Omega_s$
and $\Omega_i$. In a fibre without birefringence,
$k''_{s}=k''_{i}\equiv k''$. As a result, the TPTA after the fibre
has the same shape as TPSA, with the only difference that the
frequency arguments are replaced by scaled time arguments:
\begin{equation}
F(t_s,t_i)\propto F(\Omega_s,\Omega_i)|_{\Omega_s\equiv t_s/k''
l,\Omega_i\equiv t_i/k'' l}. \label{final}
\end{equation}

The probability of the pair to arrive at times $t_i,t_s$ is given by
the squared modulus of (\ref{final}). Distribution $|F(t_s,t_i)|^2$
can be measured directly if arrival times of the signal and idler
photons are registered separately, using, for instance, the pump
pulse as a trigger~\cite{Silberhorn}. This is not necessary for our
purposes and we measure only delays between the signal and idler
photons, in a simpler standard START-STOP technique. This way, the
information about the arrival times of signal and idler photons
separately is discarded. The situation can be explained by passing
from coordinates $t_s,t_i$ to a $45^{\circ}$ rotated frame
$t_{\pm}\equiv(t_i\pm t_s)/\sqrt{2}$. Distribution of the
coincidence counting rate versus the delay between signal and idler
arrival times, $t_s-t_i$, gives the integral of the TPTA square
modulus w.r.t. $t_+$, which corresponds, with the scaling
$\Omega_{\pm}\equiv t_{\pm}/k'' l$, to the frequency distribution of
the form
\begin{equation}
F(\Omega_-)_{\hbox{meas}}\propto
\int\hbox{d}\Omega_+|F(\Omega_+,\Omega_-)|^2, \label{marginal}
\end{equation}
where, similarly,
$\Omega_{\pm}\equiv(\Omega_i\pm\Omega_s)/\sqrt{2}$.

Distribution (\ref{marginal}) provides the projection of
$|F(\Omega_s,\Omega_i)|^2$ onto the  axis $\Omega_-$ (Fig.~1).
Additional information about TPSA can be obtained by measuring the
same distribution with narrowband filters inserted in front of the
signal or idler detectors. For infinitely narrow filters at
frequencies $\Omega_{s0}$ or $\Omega_{i0}$, the resulting
distributions will be
$\int\hbox{d}\Omega_+|F(\Omega_{s0},\Omega_i)|^2$ or
$\int\hbox{d}\Omega_+|F(\Omega_{s},\Omega_{i0})|^2$, respectively.
As a result, the signal-idler time delay distribution will show the
projection on the $\Omega_-$ axis of the TPSA intersection with the
$\Omega_i$ or $\Omega_s$ axis (shown by bold red lines in Fig.1).
The three distributions taken for the cases of (i) no filters
inserted, (ii) a filter inserted into the signal channel, and (iii)
a filter inserted into the idler channel will provide, in an easy
way, basic information about the TPSA. For the TPSA shown in Fig.1a,
these distributions are plotted in Fig. 1d. From the widths of
distributions (ii) and (iii), $\Delta\Omega_i$ and $\Delta\Omega_s$,
one can retrieve the tilt $\alpha$ of the TPSA as
$\tan\alpha=\Delta\Omega_s/\Delta\Omega_i$ (Fig.1a). From any of
these two widths and the width of the unfiltered distribution
$\Delta\Omega$, the degree of frequency entanglement~\cite{Fedorov}
can be calculated as
$R=\frac{\Delta\Omega}{\Delta\Omega_s}\frac{1}{1+\cot\alpha}=
\frac{\Delta\Omega}{\Delta\Omega_i}\frac{1}{1+\tan\alpha}$.

\begin{figure}
\includegraphics[width=1.0\textwidth]{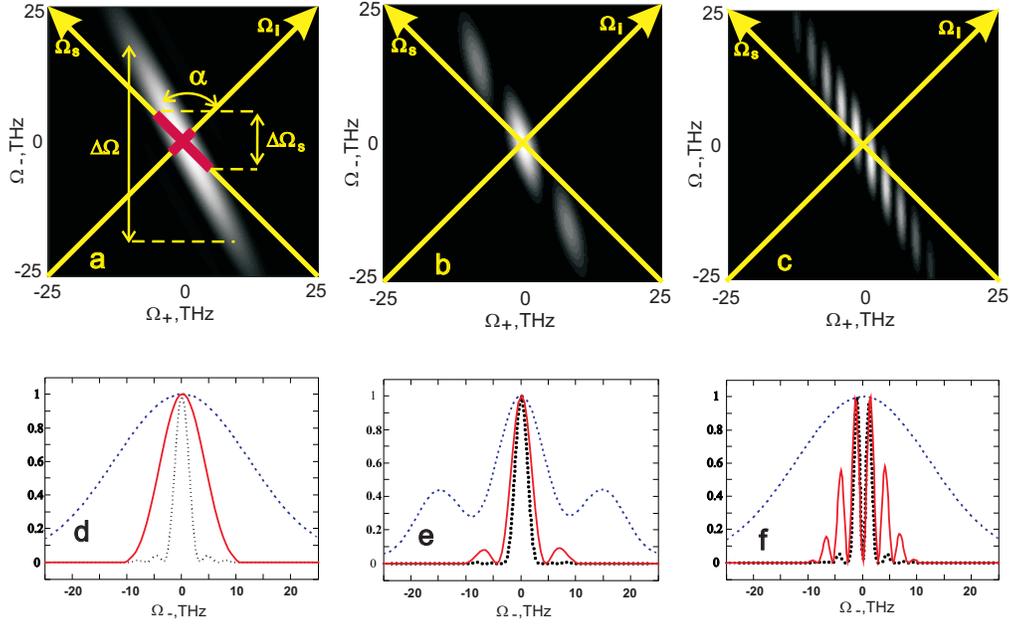}
\caption{(color online) Squared modulus of TPSA plotted in
coordinates $\Omega_{\pm}\equiv(\Omega_i\pm\Omega_s)/\sqrt{2}$
(a,b,c). The TPSA was calculated for the case of a 5 mm BBO crystal
pumped by (a) single-pulse pump and two pump pulses separated by (b)
$520$ fs and (c) $1.75$ ps. By registering coincidence counting rate
as a function of signal-idler delay, after transmitting biphotons
through an optical fibre, one retrieves the projection of TPSA onto
the $\Omega_-$ axis (figures d,e,f, blue dashed lines). If a
narrowband filter is inserted into the signal (idler) channel,
projected is not the TPSA but its cross-section along the $\Omega_i$
($\Omega_s$) axis. The corresponding distributions are shown in
figures d,e,f (black dotted lines and red solid lines,
respectively).}
\end{figure}

In other words, the three types of coincidence distributions (cases
(i), (ii), and (ii)) provide the projections of the TPSA and of its
cross-sections. These measurements also allow one to observe
interference structure in the shape of TPSA. Such structure can be
introduced deliberately, to 'engineer' the spectral properties of
biphoton light and, in particular, to get entanglement between
multiple pairs of frequency modes~\cite{psis}. The easiest way to
create a complicated shape of TPSA is to split the pump pulse in two
and hence to introduce modulation into the pump
spectrum~\cite{Kulik}. Typical shapes of TPSA in this case are shown
in Figs 1 b,c. In general, the interference structure manifests
itself in all three cases (i) (no filters), (ii) (narrowband filter
in the signal channel), and (iii) (narrowband filter in the idler
channel). At the same time, TPSA modulation with larger period
(Fig.1b) is more likely to be seen in the unfiltered distribution
than in filtered ones, since the TPSA projection will resolve the
structure (Fig.1e, blue dashed line) while its cross-section will
not (Fig.1e, black dotted line and red solid line). In the case of a
small modulation period (Fig.1c), unfiltered distribution will have
the structure 'smeared' (Fig.1f, blue dashed line) while filtering
with sufficiently small bandwidth will reveal the structure (Fig.1f,
black dotted line and red solid line).

In our experiment (Fig. 2), type-II SPDC was generated in a 5 mm BBO
crystal by means of a femtosecond-pulse pump with the wavelength 404
nm and bandwidth 2 nm. The pump was focused into the crystal by an
$f=75$ cm lens. After the crystal, the pump was eliminated using a
mirror and an RG filter, and the SPDC radiation was coupled into a
500 m of  S630 (Thorlabs) optical fibre with demagnification about
100, by means of a 20x objective lens. After the fibre, the beam was
polarization-corrected with the help of two retardation plates and
sent to a polarizing beamsplitter (PBS). Further, horizontally and
vertically polarized photons were registered, respectively, by
detectors 1 and 2 (MPD single-photon counters, jitter time $45\pm5$
ps). Distribution of the delay time between signal and idler photons
was measured by means of a time-to-amplitude converter (TAC)
followed by multi-channel analyzer (MCA).
\begin{figure}
\includegraphics[width=0.6\textwidth]{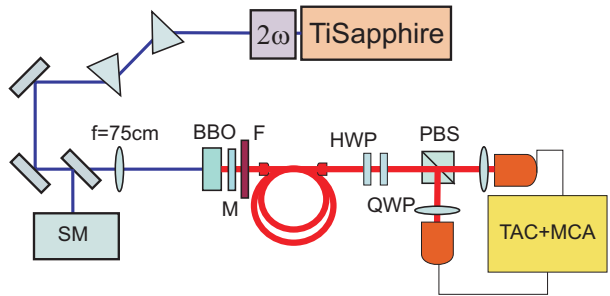}
\caption{(color online) Experimental setup. Second harmonic of a
Ti-Sapphire laser, after being cleaned from the first harmonic
radiation by two prisms, is focused into a BBO crystal. The pump
spectrum is controlled using a spectrometer (SM). After the crystal,
the pump radiation is cut off by a mirror (M) and a filter (F), and
the down-converted light is fed into 500 m of standard optical
fibre. After the fibre, polarization is corrected by means of a
half-wave plate (HWP) and a quarter-wave plate (QWP). A polarizing
beamsplitter (PBS) separates signal and idler radiation and sends
each beam to a single-photon detector. Distribution of the photon
arrival time intervals is measured by means of TAC and MCA.}
\end{figure}
The resulting time distributions for the cases of (i) no filtering,
(ii) a 1 nm filter inserted into the signal (ordinarily polarized)
beam, and (iii) a 1 nm filter inserted into the idler
(extraordinarily polarized) beam are shown in Fig.3a (respectively,
by blue triangles, black squares, and red empty circles). For
convenience, all dependencies are normalized to unity. The results
clearly demonstrate the asymmetry of TPSA with respect to signal and
idler frequencies, as the same filter inserted into signal and idler
channels leads to different time distributions. From the
experimental distributions, the tilt of the TPSA is $72\pm
0.5^{\circ}$, while from theoretical calculations (Fig.1d), its
value is $73^{\circ}$. For the degree of entanglement $R$, the
experimental distributions give a value of $1.3$, while the
theoretical distributions lead to $R=2.4$. The disagreement is
mainly caused by the finite bandwidth of the filter used ($1$ nm),
which broadens the (ii) and (iii) distributions but does not change
the (i) one, and hence reduces the $R$ value.
\begin{figure}
\includegraphics[width=0.8\textwidth]{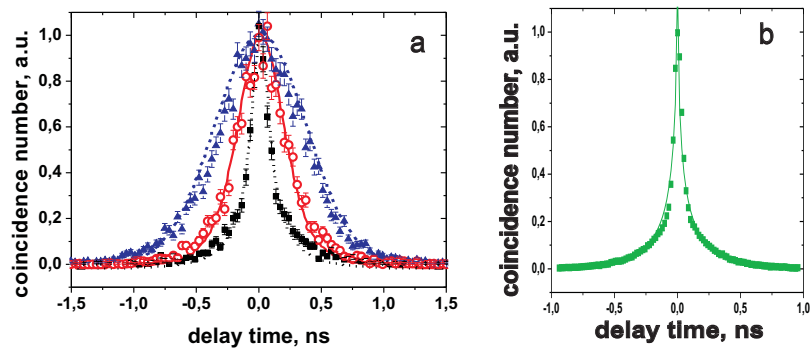}
\caption{(color online) Measured (points) and calculated (lines)
distributions of the delay time between signal and idler photons,
for the cases of no filters inserted in front of detectors (blue
triangles, blue dashed line), 1 nm filter inserted into the idler
channel (red empty circles, red solid line) and 1 nm filter inserted
into the signal channel (black squares, black dotted line). Figure b
shows the point-spread function of the method. This function was
measured by registering time delay distribution without the fibre.}
\end{figure}

Lines in Fig.3a are theoretical distributions calculated for all
three cases. Insertion of narrowband filters was taken into account
by multiplying TPSA by frequency-dependent Gaussian functions
describing the filter transmission band. When calculating time
distributions, we assumed the fibre GVD to be $k''=4.3\cdot10^{-28}
\hbox{s}^2/\hbox{cm}$. This value was found by transmitting a pulse
from the Ti-Sapphire laser first harmonic through the fibre and
measuring its autocorrelation function width. It is very close the
GVD of a standard fused silica fibre~\cite{Ranka}. The finite
resolution of the method was taken into account by registering the
coincidence distribution without the fibre inserted (Fig.3b). The
measured distribution, with FWHM equal to $90$ ps, was used as a
point-spread function, and the theoretical distributions in Fig.3a
were calculated as its convolution with the 'raw' spectra, given by
Eqs. (\ref{final},\ref{marginal}). The theoretical curves for the
filtered cases are in perfect agreement with the experimental data,
although the only fitting parameter is the height of the
distributions (all are normalized to unity). The 10\% difference in
the experimental and theoretical widths of the unfiltered
distribution can be due to the spectrum restriction by optical
elements and the detectors sensitivity curve.

In the second part of our experiment, we 'engineer' a TPSA with
modulated shape and observe this shape using our method. For this
purpose, we modify the pump spectrum using a birefringent material
with the plane of the optic axis oriented at $45^{\circ}$ with
respect to the pump polarization. This method, first suggested in
Ref.~\cite{Kulik}, allows to split the pump pulse in two temporally
separated pulses with orthogonal polarizations, oriented at $\pm
45^{\circ}$ to the initial pump polarization. The spectrum of the
pump pulse is then modulated by a harmonic function, the modulation
period scaling as the inverse time interval between the two pulses.
By introducing modulation into the pump spectrum, we also change the
TPSA (as shown in Fig.1b,c). As a birefringent material, we used a
BBO crystal of length $1$ mm and the optic axis oriented at
$45^{\circ}$ to the pump beam, which led to a $350$ fs separation
between the two pulses. Note that the crystal did not produce SPDC
radiation as its orientation did not satisfy the phase matching
conditions. After projecting the pulses onto the same polarization
direction, we obtained the pump spectrum modulated by a sine/cosine
function, depending on the phase between the pulses, which could be
aligned by slightly tilting the crystal. Fig.4 shows the pump
spectra with the modulation phase fixed at $0$ (a) and $\pi$ (b).
Further, we only focused on the case of the $\pi$ phase as the shape
is more recognizable.

\begin{figure}
\includegraphics[width=0.7\textwidth]{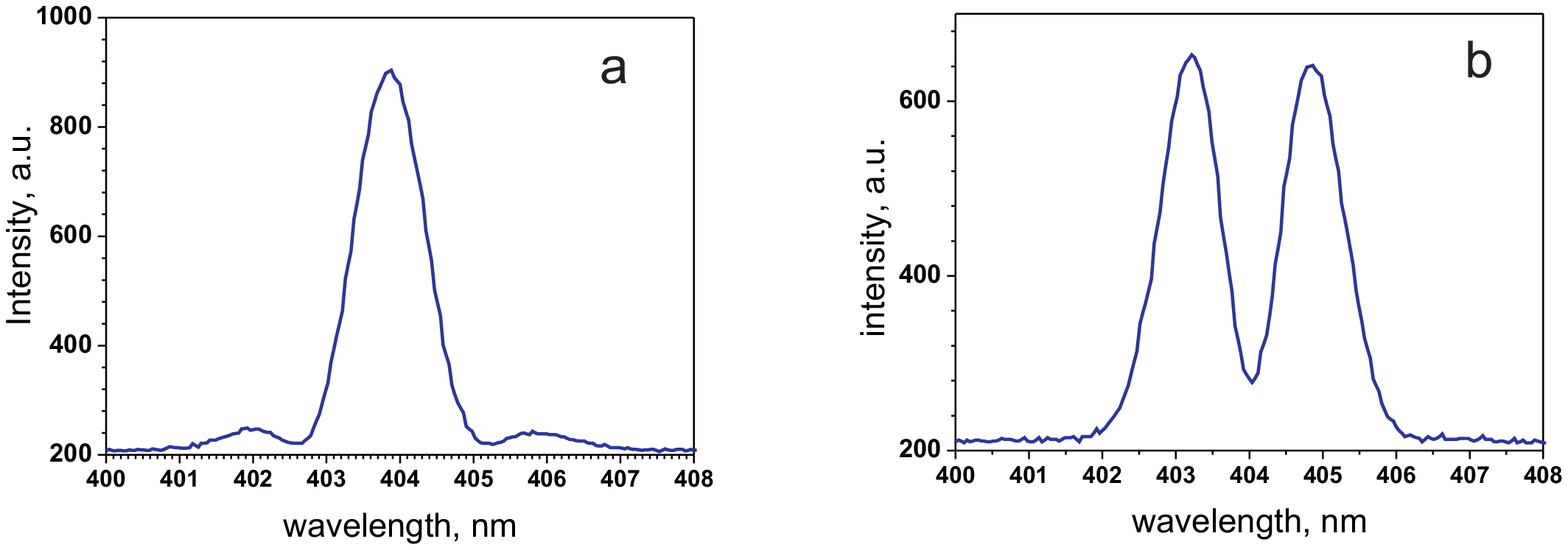}
\caption{(color online) Spectra of the pump with the modulation
created by a BBO crystal inserted into the pump beam. Left and right
figures correspond to different tilts of the crystal leading to
different phases of the modulation.}
\end{figure}

Experimental and theoretical distributions (points and lines,
respectively) are shown in Fig.5. In the case of no filters inserted
(blue triangles, blue dashed line) interference structure is clearly
seen. For filters in the signal or idler channels, we obtained no
interference structure, as predicted by theory. The plot shows the
case where a 1 nm filter is inserted into the signal channel (black
squares, black dotted line).

TPSA with the structure shown in Fig.~1c can be obtained by
inserting into the pump beam a 5-mm BBO crystal with the plane of
the optic axis oriented at $45^{\circ}$ to the vertical axis. We did
this in the experiment and indeed, the interference structure in the
unfiltered distribution was smeared, as one can predict from Figs.
1c,f. At the same time, to see the interference structure in the
filtered distributions one should perform much more narrowband
filtering than with the interference filters we used. Filtering with
$1$ nm filters did not reveal the interference structure.

\begin{figure}
\includegraphics[width=0.6\textwidth]{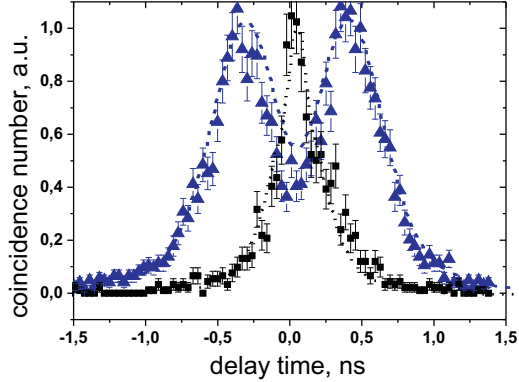}
\caption{(color online) Measured (points) and calculated (lines)
distributions in the case of the pump spectrum shown in Fig. 4b.
Blue dashed line and blue triangles correspond to the unfiltered
case, black dotted line and black squares, to the case of a 1 nm
filter inserted into the signal channel.}
\end{figure}

In conclusion, we have performed an experiment aimed to demonstrate
a technique for studying the TPSA of ultrafast (femtosecond-pulsed)
two-photon states. By measuring the distribution of the time delay
between signal and idler photons, we are able to retrieve the
projection of TPSA on the $\Omega_+$ axis. With the help of
narrowband filters inserted into signal and idler channels, one can
measure the tilt of TPSA and the degree of entanglement. Further, we
used this technique for showing the effect on the shape of TPSA of
the pump spectrum modulation. This last result paves the way, in
view of application to quantum technologies\cite{mg}, for
engineering the biphoton TPSA, as, for example, the generation of
controlled optical frequency combs \cite{psis}.

We are grateful to L.~A.~Krivitsky for the help at the early stages
of the experiment. This work has been supported in part by MIUR
(PRIN 2007FYETBY), "San Paolo foundation", NATO (CBP.NR.NRCL
983251), RFBR 08-02-00555a and the Russian Program for Scientific
Schools Support, grant \# NSh-796.2008.2. M.~V.~Ch. also
acknowledges the support of the CRT Foundation.


\begin{thebibliography}{99}

\bibitem{Grice}W.P.~Grice and I.A.~Walmsley, Phys. Rev. A  \textbf{56}, 1627 (1997).

\bibitem{Keller}T.E.~Keller and M.H.~Rubin, Phys. Rev. A \textbf{56}, 1534-1541 (1997).

\bibitem{Fedorov} M.~V.~Fedorov, M.~A.~Efremov, A.~E.~Kazakov, K.~W.~Chan, C.~K.~Law, and J.~H.~Eberly, Phys. Rev. A
\textbf{69}, 052117 (2004); Yu.M.~Mikhailova, P.A.~Volkov,
M.V.~Fedorov, Phys. Rev. A \textbf{78}, 062327 (2008); P.~A.~Volkov,
Yu.~M.~Mikhailova, M.~V.~Fedorov, Advanced Science Letters
\textbf{2}, 511 (2009); G.~Brida, V.~Caricato, M.~V.~Fedorov,
M.~Genovese, M.~Gramegna and S.~P.~Kulik, EPL \textbf{87} 64003
(2009).

\bibitem{Mosley} P.~J.~Mosley, J.~S.~Lundeen, B.~J.~Smith, P.~Wasylczyk,
A.~B.~U'Ren, C.~Silberhorn, and I.~A.~Walmsley, Phys. Rev. Lett.
\textbf{100},  133601 (2008).

\bibitem{Kim&Grice} Y.-H.~Kim  and W.~P.~Grice,
Opt. Lett. \textbf{30}, 908 (2005).

\bibitem{Wasilewsky} W.~Wasilewski, P.~Wasylczyk, P.~Kelenderski, K.~Banasek, and
C.~Radzewicz, Opt. Lett. \textbf{31}, 1130 (2006).

\bibitem{Poh} H.~S.~Poh, C.~Y.~Lum, I.~Marcikic, A.~Lamas-Linares, and C.~Kurtsiefer,
Phys. Rev. A \textbf{75}, 043816 (2007); X. Shi et al., Optics
Letters \textbf{33}, 875 (2007); A.~Valencia, A.~Cere, X.~Shi,
G.~Molina-Terriza, and J.~P.~Torres, PRL \textbf{99}, 243601 (2007);
M.~Hendrych, X.~Shi, A.~Valencia, and J.~P.~Torres, Phys. Rev. A
\textbf{79}, 023817 (2009).

\bibitem{spectronPRL} A.~Valencia, M.~V.~Chekhova, A.~S.~Trifonov and Y.~H.~Shih,
Phys. Rev. Lett. \textbf{88},  183601 (2002).

\bibitem{spectron} M.~V.~Chekhova, JETP Lett., \textbf{75}, 225-226 (2002).

\bibitem{BellPRL} G.~Brida, M.~V.~Chekhova, M.~Genovese, M.~Gramegna, and L.~A.~Krivitsky,
Phys. Rev. Lett. \textbf{96},  143601 (2006); G.~Brida,
M.~V.~Chekhova, M.~Genovese and L.~Krivitsky, Phys. Rev. A
\textbf{76}, 053807 (2007).

\bibitem{Silberhorn} M.~Avenhaus, A.~Eckstein, P.~J.~Mosley,  and C.~Silberhorn,
Opt. Lett. \textbf{34}, 2873 (2009).

\bibitem{Kim} S.~Y.~Baek, O.~Kwon, and Y.-H.~Kim, Phys. Rev. A \textbf{78}, 013816 (2008).

\bibitem{psis} N.~C.~Menicucci, S.~T.~Flammia, O.~Pfister, \PRA \textbf{76} 010302 (2007); \PRL
\textbf{101} 130501 (2008).

\bibitem{Kulik} Y.-H.~Kim, M.~V.~Chekhova, S.~P.~Kulik, Y.-H.~Shih, and
M.~H.~Rubin, Phys. Rev. A \textbf{61}, 051803(R) (2000).

\bibitem{Ranka} J.~K.~Ranka, R.~S.~Windeler, and A.~J.~Stentz, Opt.
Lett.  \textbf{25}, 25 (2000).

\bibitem{mg} M. Genovese, Phys. Rep. \textbf{413}, 3197 (2005).




\end{thebibliography}
\end{document}